\documentclass[prd,aps,twocolumn,nofootinbib]{revtex4}
\usepackage{graphicx}
\usepackage{color}
\usepackage{amssymb}
\def\gr{$\gamma$-ray}

\begin{document}
\title{Cosmic ray composition measurements and cosmic ray background free  \gr\ observations  with Cherenkov telescopes}
\author{Andrii Neronov}
\affiliation{Astronomy Department, University of Geneva, Ch. d'Ecogia 16, 1290, Versoix, Switzerland}
\author{Dmitri V. Semikoz}
\affiliation{APC, Universite Paris Diderot, CNRS/IN2P3, CEA/IRFU,\\
Observatoire de Paris, Sorbonne Paris Cite, 119 75205 Paris, France and\\
National Research Nuclear University MEPHI (Moscow Engineering Physics Institute), Kashirskoe highway 31, Moscow, 115409, Russia}
\author{Ievgen Vovk and Razmik Mirzoyan}
\affiliation{Max Planck Insitut fur Physik, Foehringer Ring 6, 80805, Munich, Germany}

\begin{abstract}
Muon component of extensive air showers (EAS) initiated by cosmic ray particles carries information on the primary particle identity. We show that the muon content of EAS could be measured in a broad energy range from 10-100 TeV up to ultra-high-energy cosmic ray range using  wide field-of-view imaging atmospheric Cherenkov telescopes observing strongly inclined or nearly horizontal EAS from the ground of from high altitude. Cherenkov emission from muons in such EAS forms a distinct component  (halo or tail) of the EAS image in the telescope camera. We show that detection of the muon signal could be used to measure composition of the cosmic ray spectrum in the energy ranges of the knee, the ankle and of the Galactic-to-extragalactic transition. It could also be used to veto the cosmic ray background in gamma-ray observations. This technique provides a possibility for up to two orders of magnitude improvement of sensitivity for \gr\ flux in the energy band above 10 PeV, compared to KASCADE-Grande, and an order-of-magnitude improvement of sensitivity in the multi-EeV energy band, compared to  Pierre Auger Observatory.  
\end{abstract}
\maketitle

\section{Introduction}

Measurement of identity of the primary cosmic ray particles with energies above PeV is usually based on the  information on the particle content of Extensive Air Showers (EAS).  EAS-TOP, CASA-MIA, KASCADE and KASCADE-Grande \cite{EAS-Top,CASA-MIA,kascade05,kascade,kg15} experiments have constrained the composition of the cosmic ray flux in this energy range   via a combined measurement of the muon and electron signals of the EAS reaching the ground level.
EAS initiated by heavier nuclei tend to be systematically more muon-rich. Pierre Auger Observatory \cite{auger_composition,auger_composition1} has derived the cosmic ray flux composition in the energy range above 100 PeV  from  the measurement of the longitudinal profiles of the EAS development in the atmosphere.  EAS initiated by heavier nuclei tend to reach maximum at shallower depths in the atmosphere. Auger Prime upgrade of Pierre Auger Observatory aims at modification of surface detector which will separately measure the electromagnetic and muon content of EAS initiated by Ultra-High-Energy Cosmic Rays (UHECR), thus enabling a combinaiton of the measurement of the depth of the shower maximum with the measurements of the muon content of the EAS to improve the measurement of the UHECR particle identity \cite{augerprime}. 

Measurement of the composition of the cosmic ray flux in the PeV-EeV energy range provides a clue to the problems of the origin of the knee and ankle of the cosmic ray spectrum and of the energy of transition between Galactic and extragalactic cosmic ray flux \cite{giacinti15,giacinti15a,unger15,globus15}. KASCADE data suggest that features in the cosmic ray spectrum are produced by a sequence of softenings of the spectra of heavier elements at higher energies \cite{kascade11}. In the energy range above few times $10^{17}$~eV, beyond the energy of the knee of the iron nuclei, the composition of the spectrum becomes light \cite{LOFAR,tunka}. This fact, combined with the moderate anisotropy of the cosmic ray flux might indicate the onset of extragalactic component of the  flux in this energy range \cite{giacinti15,giacinti15a}. 

Interpretation of the data on cosmic ray flux composition is strongly dependent on hadronic interaction models. Model uncertainties introduce a systematic error in the compositon measurements. Better control of this error could be achieved via a measurement of additional EAS parameters \cite{kascade_hadronic,kascade}.  

Measurement of the muon content and of the depth of the shower maximum are  also useful for gamma-ray astronomy because they allow to distinguish the EAS initiated by primary \gr s from those initiated by protons and heavy nuclei. Gamma-ray flux  from astronomical sources emitting in the energy range above TeV is typically much weaker than the cosmic ray flux from the same direction on the sky. The sensitivity of \gr\ telescopes is limited by the capabilities of rejection of the cosmic ray background.  Imaging Atmospheric Cherenkov Telescopes (IACTs) sensitive in the multi-TeV energy range currently reach the background suppression down to $\sim 10^{-2}$ of the cosmic ray flux \cite{cta_montecarlo}, using a technique of analysis of the shapes of the EAS images in the telescope camera. 

 The \gr\ induced showers are muon-poor  \cite{engel}. KASCADE and KASCADE-Grande experiments have used this fact to study \gr\ luminosity of the Northern sky in the energy range above 100 TeV, reaching the charged cosmic ray background rejection factor $\sim 10^5$, based on the detection of muons in the EAS \cite{kascade_ICRC}.  No \gr s were detected in this energy band. However, IceCube observation of the astrophysical neutrino signal in the same energy band \cite{IceCube_PeV,IceCube_3yr,IceCube_ICRC} might be suggesting that the Milky Way is producing detectable  0.1-1~PeV \gr\ and neutrino emission \cite{neronov14a,neronov15,neronov16}.  The inner Galaxy, from which the \gr\ flux is stronger \cite{ahlers_murase,neronov16}, is not visible from the Northern hemisphere observed by KASCADE. \gr\ observations reaching high cosmic ray background rejection level comparable to that attained by KASCADE, but in the Southern hemisphere, are desirable. 

The KASCADE / KASCADE-Grande limit relaxes above  10 PeV. At still higher energies, above $10^{18}$~eV, limits on the \gr\ luminosity of the sky at the level of several percent of the cosmic ray flux  are imposed by  Pierre Auger Observatory \cite{auger_gamma,auger15} which uses the measurements of the depth of the shower maximum to separate \gr\ induced EAS  from proton and heavier nuclei EAS. 

In what follows we show that IACT systems potentially have a capability to measure the cosmic ray / \gr\ primary particle identities. We show that they could measure the muon content of the EAS and thus adopt the KASCADE approach for the measurement of the cosmic ray composition. Muons, similarly to electrons and positrons in the EAS, produce detectable Cherenkov light. Muons are less abundant, compared to electrons and positrons, in the depth range close to the shower maximum. However, muons propagate over much larger distances. Larger length of muon tracks increases the cumulative Cherenkov photon yield for muons in  strongly inclined or horizontal EAS. We also show that measurements of the muon content of the EAS could be combined with the measurements of the depth of the shower maximum, also possible with the IACT systems.  A combined measurement of the Cherenkov signal from electron and muon  components of strongly inclined showers and of the depth of the shower maximum provides a powerful tool for the study of the elemental composition of the cosmic ray flux and for the rejection of the cosmic ray background in gamma-ray observations.

\section{Cherenkov emission from muon and electron components of the EAS}

\begin{figure}
\includegraphics[width=\linewidth]{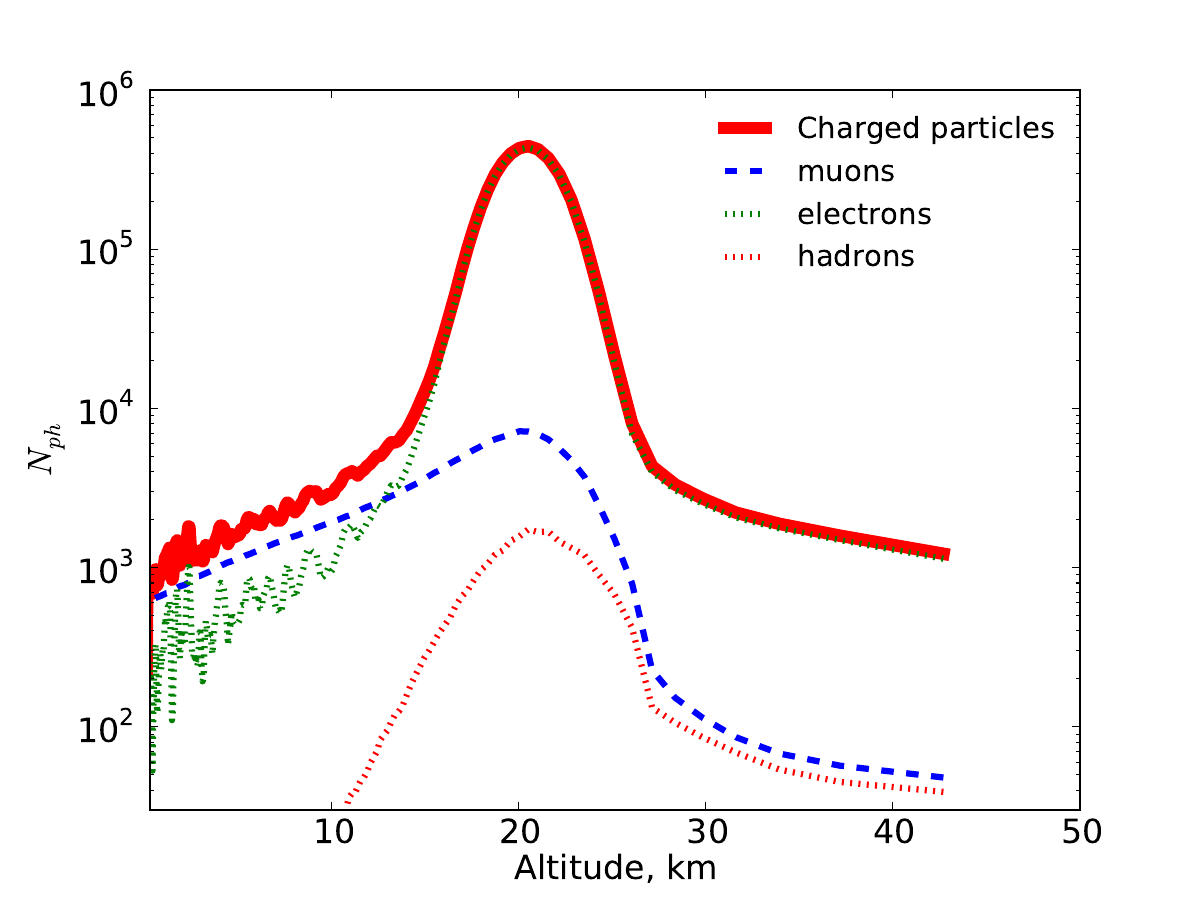}
\caption{Longitudinal profiles of charged particle distribution in an EAS initiated by a 1 PeV energy proton inclined at $87^\circ$.}
\label{fig:profile_particles} 
\end{figure}

Cherenkov emission is produced by charged  particles (mostly electrons/positrons and muons) with gamma factors larger than $\gamma_{Ch}\simeq 1/\sqrt{n^2-1}\simeq 40$, where $n\simeq 1.0003$ is the refraction index of the air (in the Troposphere). Electrons with gamma factors $\gamma\sim \gamma_{Ch}$ loose their energy mostly via the ionisation loss on rather short distance scale $l_{e,ion}\sim 10^2\left(\gamma/\gamma_{Ch}\right)(\rho_{air}/\rho_0)^{-1}$~m in the densest part of the atmosphere with the near-ground density $\rho_{air}\sim \rho_0\simeq 1.2$~g/cm$^3$. Higher energy electrons with gamma factors $\gamma\gtrsim 10\gamma_{Ch}$ loose energy predominantly onto Bremsstrahlung on the radiation length  distance scale $l_{e,Br}\sim 10^3(\rho_{air}/\rho_0)^{-1}$~m.  The electron component of proton induced  EAS  develops on the distance scale of about $\sim 5 (\rho_{air}/\rho_0)^{-1}$~km for the PeV energy scale showers (see Fig. \ref{fig:profile_particles}, calculaiton for the figure is done using CORSIKA air shower simulation program\footnote{https://www.ikp.kit.edu/corsika/ version 740 with high energy hadronic interactions model QGSJET II-03 and low energy hadronic interactions model UrQMD 1.3.1}). Beyond this distance, the number of electrons capable to emit Cherenkov light drops. 

\begin{figure}
\includegraphics[width=\linewidth]{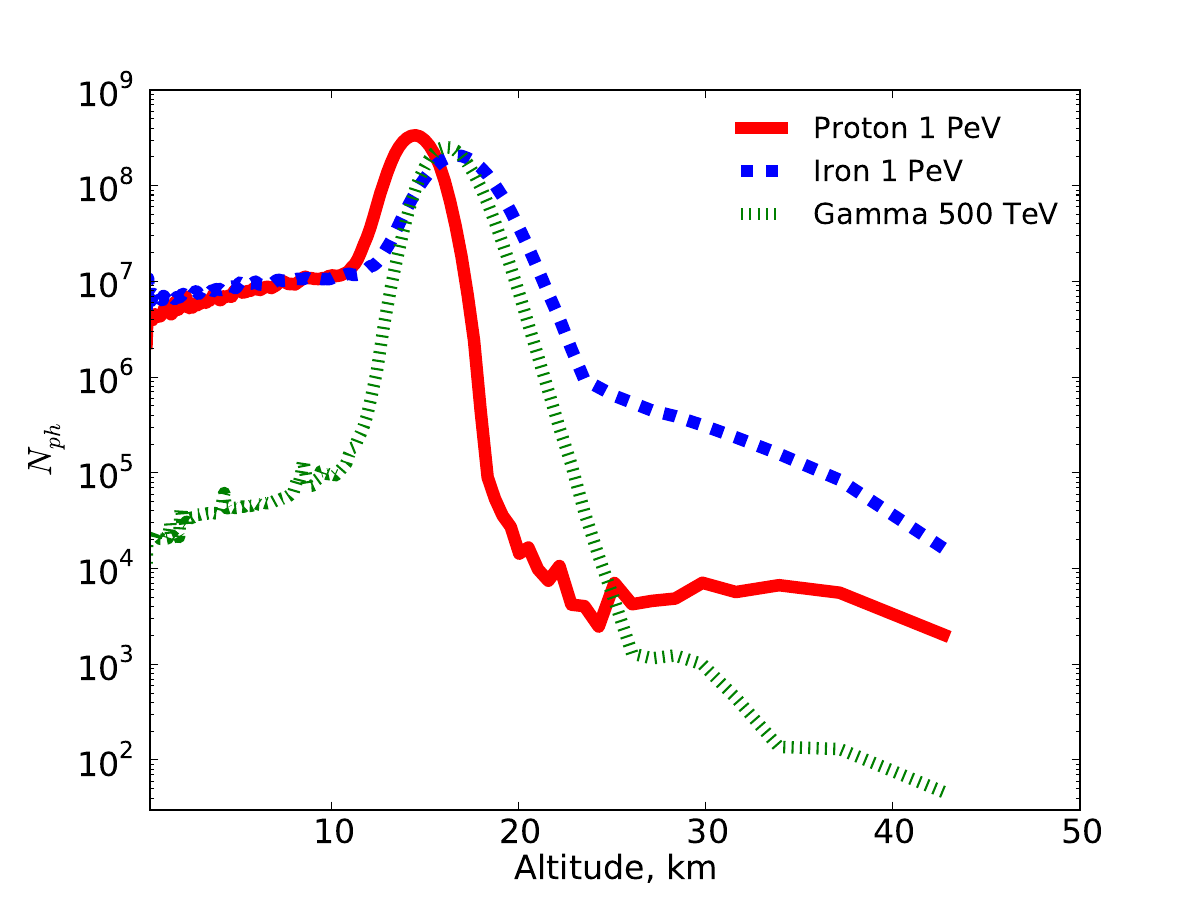}
\caption{Longitudinal profiles of Cherenkov light emission by the EAS initiated by proton (red solid line) iron nucleus (blue dashed line) and by \gr\ (green dotted line) incident at $\theta_z=87^\circ$.}
\label{fig:long_profiles} 
\end{figure}

Contrary to electrons, muons emitting Cherenkov photons suffer from much lower energy loss. Their ionisation loss distance -- $l_{\mu,ion}\sim 2\times 10^4\left(\gamma/\gamma_{Ch}\right)(\rho_{air}/\rho_0)^{-1}$~m -- is comparable to the decay length of muons $\lambda_\mu=2\times 10^4\left(\gamma/\gamma_{Ch}\right)$~m in the densest part of the atmosphere (for the low energy muons with $\gamma\sim\gamma_{Ch}$). Bremsstrahlung and pair production energy losses become important only on the distance scale $l_{\mu, Br}\sim 5\times 10^5(\rho_{air}/\rho_0)^{-1}$~m for muons with the energies above $\sim 100$~GeV \cite{pdg}.

The difference in the energy losses of electrons and muons explains the change in the particle content of the shower with the decrease of the altitude, shown in Fig.~\ref{fig:profile_particles}. In an EAS initiated by a 1~PeV proton incident at zenith angle $\theta_z=87^\circ$ electrons dominate the EAS in the altitude range above $h_{e\mu}\simeq 10$~km, while muons dominate the shower below this altitude. 
\begin{figure}
\includegraphics[width=\linewidth]{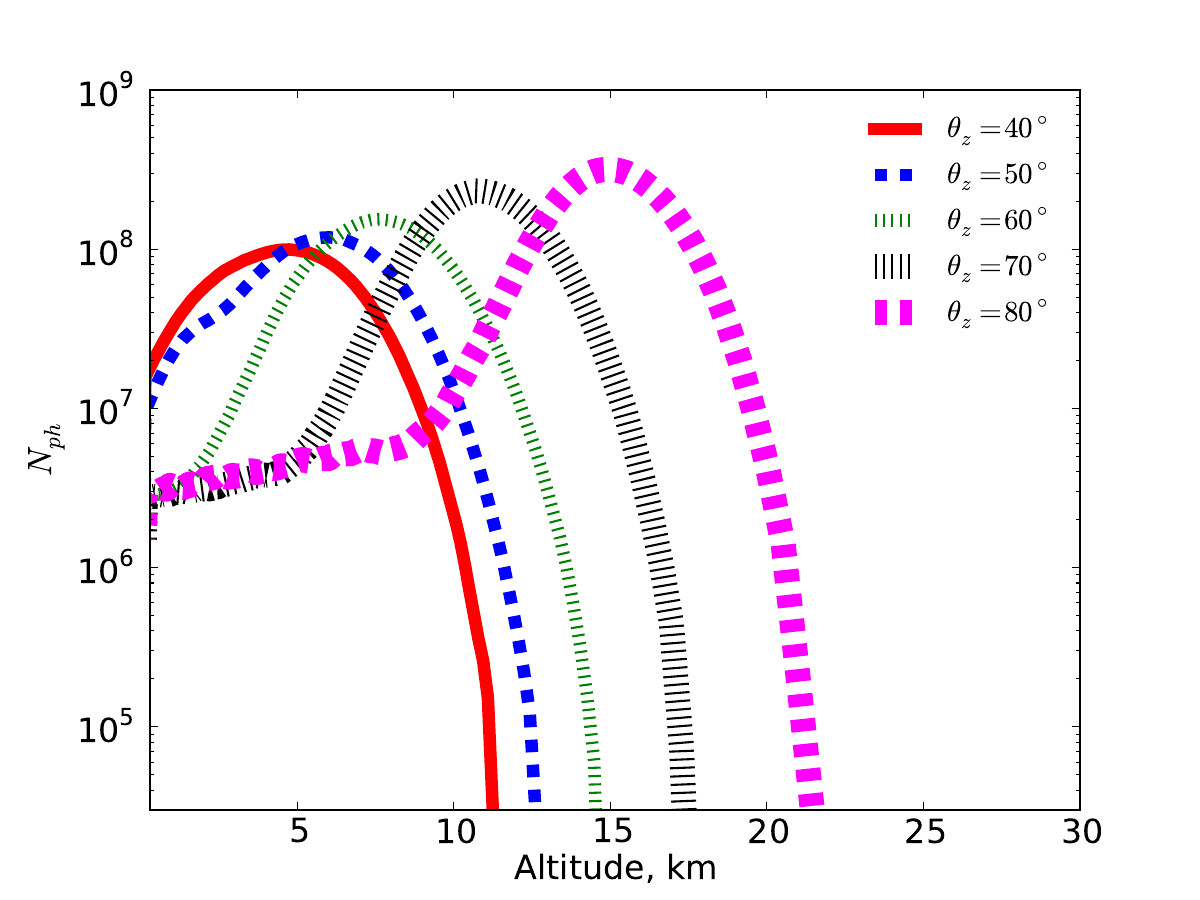}
\caption{Longitudinal profiles of the Cherenkov light emission from the EAS initiated by 1 PeV protons incident at different zenith angles.}
\label{fig:profiles_thetaz} 
\end{figure}

Change of the particle content of the shower also explains the shape of the longitudinal intensity profile of the Cherenkov light, shown in Fig.~\ref{fig:long_profiles}.  The bump visible in the altitude range 10-20~km is attributed to the Cherenkov emission from electrons and the ``plateau'' in the altitude range $h<h_{e\mu}$ is due to the Cherenkov emission from muons.  

Fig.~\ref{fig:long_profiles} also provides a comparison of the longitudinal profiles of Cherenkov light from the EAS generated by different primary particles: protons, iron nuclei and \gr s. Qualitative difference between the profiles is readily understood. 

The iron induced shower has the same ``bump plus plateau'' shape, due to the electron and muon components. However, the electron ``bump'' is shifted toward the smaller depth. This is due to the fact that the iron shower is split onto a large number of lower energy nucleon sub-showers through the spallation at the top of the atmosphere. The iron induced showers are also more muon-rich, as it is clear from the higher relative  normalisation of the muon ``plateau'' in the iron-induced shower profile (Fig.~\ref{fig:long_profiles}). These muons are predominantly generated by the decays of charged pions produced in nucleon interactions and such production channel is absent in the \gr\ induced showers. Thus, the qualitative difference of the EAS initiated by a \gr\ is the absence of the muon ``plateau''.

\section{Imaging of Cherenkov light from muon and electron components of EAS}

The difference in the  longitudinal profiles of Cherenkov emission between protons, heavier nuclei and \gr s could in principle be measured by the Cherenkov telescopes. However, this is hardly possible for the showers incident at low zenith angles.  In such showers the altitude at which the muon component starts to dominate the shower particle content is too low (or even below the ground level), see Fig. \ref{fig:profiles_thetaz}.  This way the muon plateau appears only in the showers inclined by $\theta_z\gtrsim 70^\circ$.

The electromagnetic component of a strongly inclined shower develops above the top of the Troposphere at large distances from the observer, in the distance range $r\gtrsim H_{atm}/\cos\theta_z$, where $H_{atm}\simeq 8$~km is the scale height of the atmosphere. The distance $r$ reaches the maximum for the horizontal EAS:
\begin{equation}
r\gtrsim r_{horizontal}=\sqrt{2H_{atm}R_\oplus}\simeq 320\mbox{ km}
\end{equation}
where $R_\oplus=6375$~km is the Earth radius. Cherenkov emission from electrons, however, dominates only over a small fraction of the distance range $r$. The larger part of the length range  contains mostly the Cherenkov light from the muon component of the shower.  In spite of the lower intensity of the Cherenkov emission from the muon component  (by a factor of $\sim 10-30$ for proton showers, see Fig. \ref{fig:long_profiles}), the cumulative number of Cherenkov photons generated by muons might be comparable to that originating from the electron component in the case of nearly horizontal showers. Besides, the muon induced Cherenkov light is emitted closer to the ground. This boosts the flux of the muon Cherenkov emission, compared to that of electrons. 

\begin{figure}
\includegraphics[width=\linewidth]{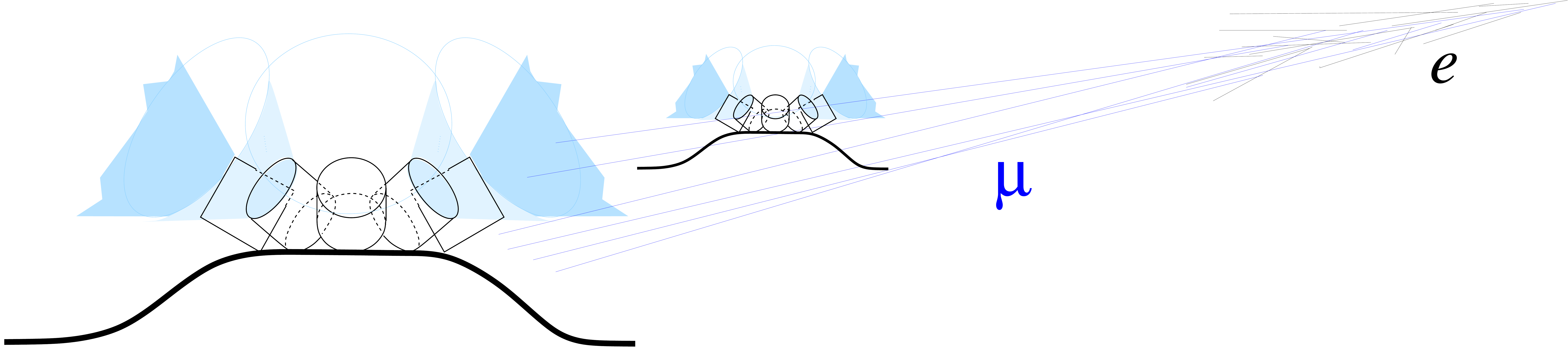}
\caption{
Principle of observations of muon and electron components of nearly horizontal EAS with wide FoV IACTs overlooking the horizon strip. Each telescope module is shown with a black cylinder. The FoVs of individual telescope modules are shown by the light blue cones. Telescopes are grouped in two stations to provide stereo view of the EAS. 
}
\label{fig:scheme} 
\end{figure}

A setup suitable for the observations of the Cherenkov light from the muon component of the EAS should consist of the Cherenkov telescopes overlooking a strip around the Earth horizon from a high altitude (e.g. the top of a mountain of the altitude  $\gtrsim 1.5$~km (typical scale height of the inversion layer), to avoid the aerosol layer above the Earth surface), as it is shown in   Fig. \ref{fig:scheme}. Largest exposure could be achieved with a moderately large set of telescopes with a wide (up to $\Theta_{FoV}=60^\circ$) Field of View (FoV) optics like e.g. the optics of the EUSO \cite{euso} or CHANT \cite{chant}  telescopes, overlooking the entire $360^\circ$ strip along the horizon. The overall arrangement of the telescopes could be similar to that implemented in the fluorescence telescopes of the Pierre Auger Observatory and Telescope Array experiments. The main difference in this case would be optimisation of the telescopes for detection of Cherenkov, rather than fluorescence, signal from the EAS.
Alternatively, a very large number of telescopes with moderate FoV (e.g. with $\Theta_{FoV}\sim 9^\circ$ like the Small Size Telescopes (SST) of the CTA \cite{CTA_design}) might be specially configured for the high zenith angle observations, to provide an overview of the entire $360^\circ$ strip above the horizon, with each of the 50-70 SST units pointing in a different direction.

\subsection{Longitudinal profiles of nearly horizontal showers in the telescope camera}

\begin{figure}
\includegraphics[width=\linewidth]{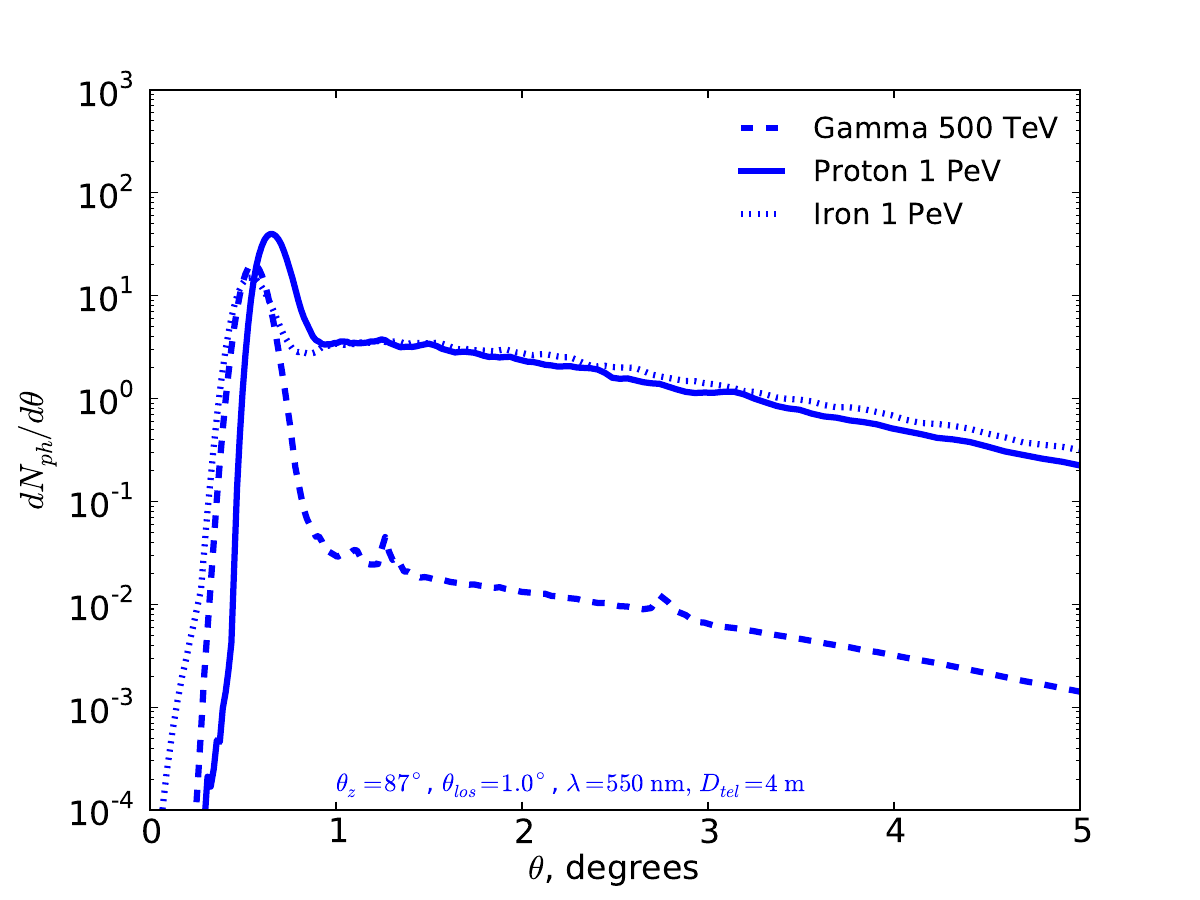}
\includegraphics[width=\linewidth]{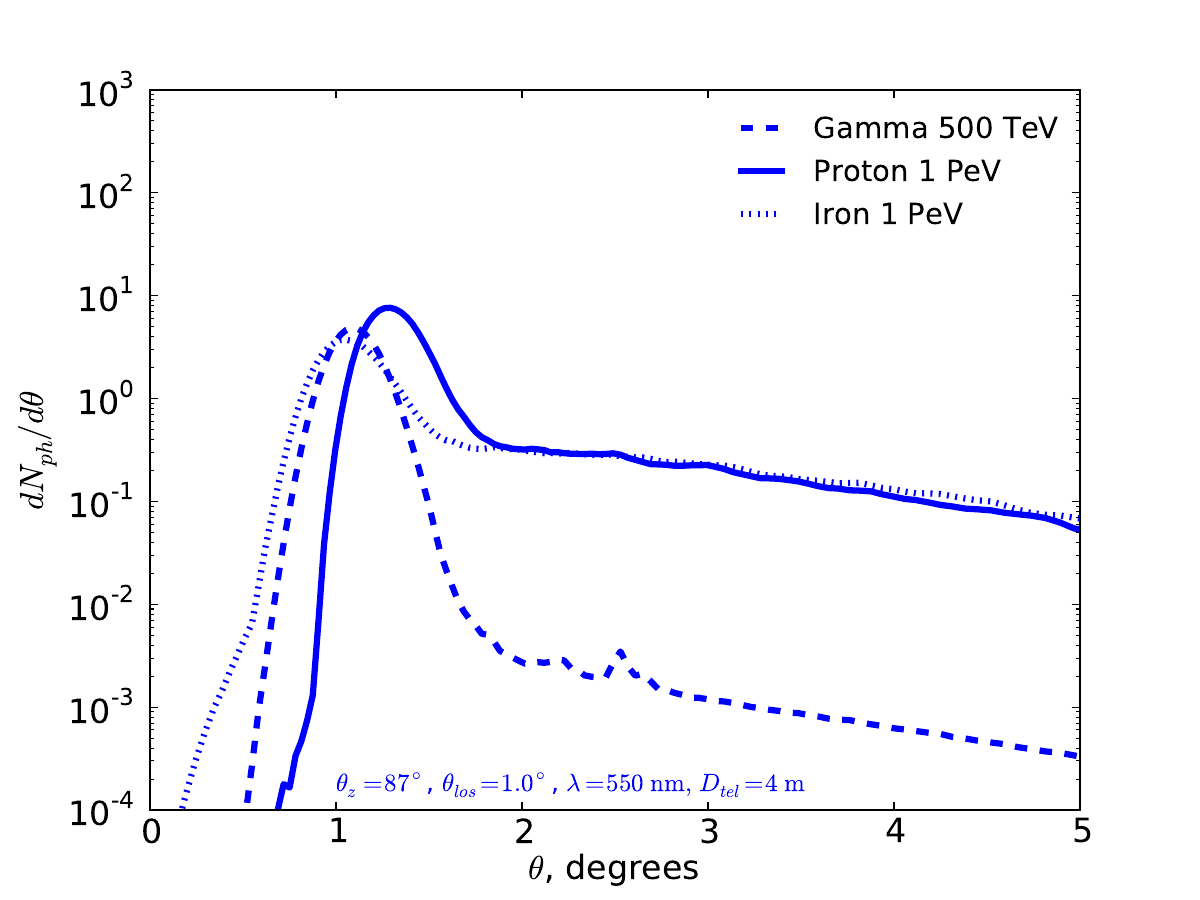}
\caption{Longitudinal profiles of proton, iron and \gr\  EAS images in the telescope camera for the nearly horizontal shower direction. The two panels correspond to the different shower misalignment with the line of sight: $\theta_{los}=0.5^\circ$ is assumed in the top and $\theta_{los}=1^\circ$ in the bottom one.}
\label{fig:profile_cherenkov} 
\end{figure}

From Fig.~\ref{fig:long_profiles} one could see that the electron  component of a nearly horizontal EAS develops in an atmospheric layer above the Troposphere, in the altitude range $h_e\simeq 10-20$~km. Lower density of the air in this high altitude layer makes the electromagnetic shower long, with the typical length scale $l\sim H_{atm}\cdot \exp(h_e/H_{atm})\sim 10^2$~km. The angular scale of this component in the telescope camera is 
\begin{equation}
\theta_{em}\sim\theta_{obs} \frac{l}{r_{\rm horizontal}}\sim 0.3^\circ
\end{equation}
where $\theta_{obs}\lesssim 1^\circ$ is the observation angle (the angle between the shower axis and the direction toward the telescope). The electromagnetic cascade part of the EAS is only marginally resolvable with IACT systems which typically have the Point Spread Function (PSF) of about $0.1^\circ$ (the actual pixel size of the IACT cameras could be larger). 

Fig.~\ref{fig:profile_cherenkov} shows the longitudinal profiles of the EAS from Fig. \ref{fig:long_profiles}, as they would appear in a camera of a $D_{tel}=4$~m diameter IACT pointed in the horizontal direction. The efficiency of the telescope optical system (product of the throughput of the optical system and the photon detection efficiency of photosensors)  is assumed to be $\epsilon=0.1$. The shower axis is assumed to be misaligned with the line of sight by $\theta_{los}=0.5^\circ$ (top) or $\theta_{los}=1^\circ$ (bottom). This corresponds to the shower impact parameters $d_{EAS}\sim r_{horizontal}\theta_{obs}\simeq 2.5$~km and $5$~km. 

The atmosphere is not transparent to the UV light in the horizontal direction. Taking this into account, we have considered photosensors sensitive in the visible band  (reference  wavelength $\lambda=550$~nm) for the calculation of Fig.~\ref{fig:profile_cherenkov}. Observations in the visible band lead to a decrease of the Cherenkov signal, which has the spectrum $dN/d\lambda\propto \lambda^{-2}$.  Also, a possible aerosol absorption will play a significant role. This calculation takes into account the Rayleigh scattering with the characteristic $\lambda^4$ scaling of the scattering length.  It also takes into account the distance-square scaling of the Cherenkov photon flux and the exponential angular distribution of the Cherenkov photons around the EAS axis (from Ref.~\cite{nerling}). 

Comparing Fig.~\ref{fig:profile_cherenkov} with Fig.~\ref{fig:long_profiles} one can identify the muon component contribution to the EAS image as a long multi-degree-scale extension of the shower image profile beyond the compact electron ``bump''.

\subsection{Formation of the muon tail of the shower images}

Observational appearance of the  muon component of $\theta_z \gtrsim 70^\circ$ EAS images is different  from that of the  vertical showers. The vertical showers develop in much thinner atmosphere at the distance of $\sim 10$~km from the telescope. The presence of a sub-dominant muon component of the shower (in terms of the number of particles) could be noticed only if the individual muons produce parts of the so-called muon rings arriving simultaneously with the Cherenkov signal from the electron component of the shower.  The muon ring signal could be strong enough for the muons passing close to the telescope and producing Cherenkov light with the density $\sim 10 \mathrm{ph/m^2}$ on the ground~\citep{Mirzoyan_single_muons} -- sufficient to be triggered by telescope camera.

The strongly inclined showers develop at much larger distance ($\sim 30-300$~km at $\theta_z \gtrsim 70^\circ$). The density of muons reaching the ground is diluted by this large distance scale and by the decay of the lower energy muons. This reduces the probability for a single muon to produce a significant contribution to the EAS image. 

\begin{figure}
  \includegraphics[width=0.9\linewidth]{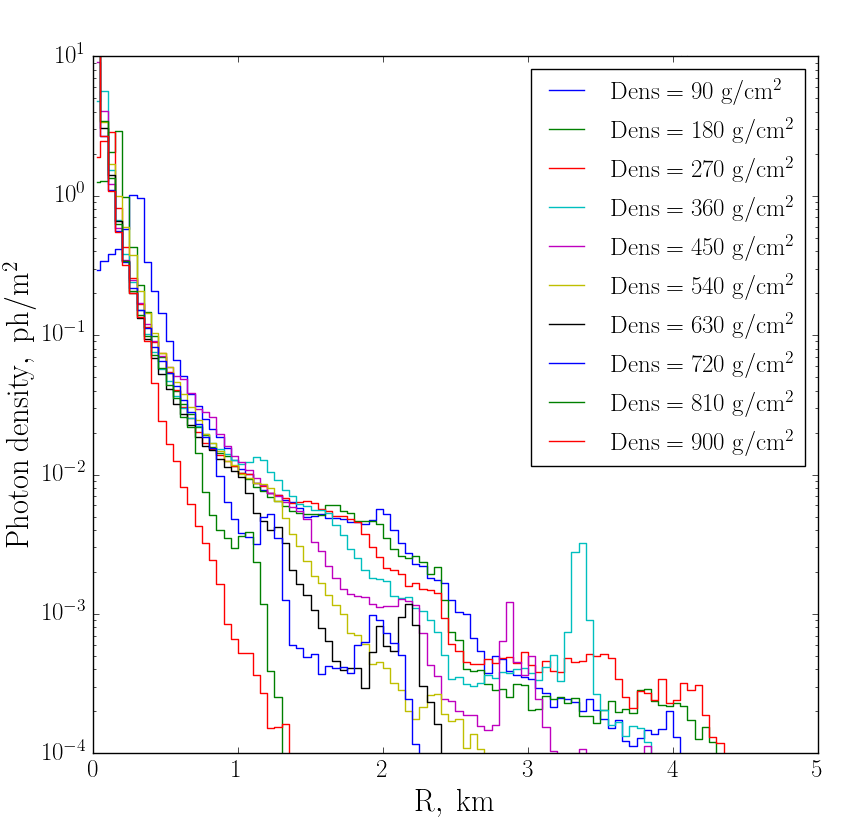}
  \caption{Radial density profile of the Cherenkov photons, emitted by a single 1~TeV muon starting at different heights in the atmosphere. The latter is parametrised with the atmosphere vertical column density, where zero corresponds to the infinite height.}
  \label{fig:muon_density_profile} 
\end{figure}

Fig.~\ref{fig:muon_density_profile} shows the radial density profile of the photons reaching the ground from a single muon emitted at $\theta_z = 87^\circ$. One could see that the muon light pool extends up to 2-3~km from the muon direction axis compared to just $R_\mu^{vert}\simeq 10^2$~m in the vertical case~\citep{Mirzoyan_single_muons}. Since the Cherenkov angle does not significantly change between these cases and stays $\alpha_{Ch} \simeq 1^\circ ... 1.5^\circ$, this difference is the result of the simple geometrical scaling of the light pool to the larger distances from the telescope:
\begin{equation}
  \label{eq:muon_R_scaling}
  R_{\mu} \simeq R_{\mu}^{vert} \frac{r_{horizontal}} {r_{vertical}} \simeq 3.5 \mbox{ km}
\end{equation}

\begin{figure}
  \includegraphics[width=\linewidth]{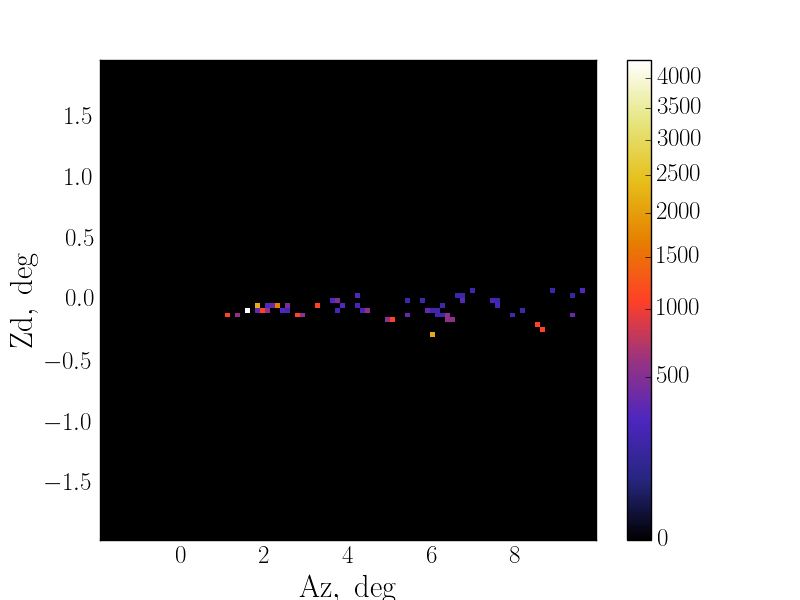}
  \caption{Simulated image of the muon component of a 1~TeV proton shower, constructed from the single muon images from different heights in the atmosphere. The images were added with the weights corresponding to the muons density at the appropriate heights, estimated from the simulation (see Fig.~\ref{fig:profile_particles}). The image was produced for the telescope with $d=50$~m dish and corresponds to the impact parameter $d_{EAS}=4$~km.}
  \label{fig:simulated_proton_image} 
\end{figure}

The typical Cherenkov photon density produced by a single muon at a several kilometre distance is  $\lesssim 10^{-2}~\mathrm{ph/m^2}$.  Single muons can no longer be detected by the telescope. However, the collective signal from a large number of muons (say, $\sim 10^4$) within the several kilometre large muon Cherenkov light pool is detectable. To directly verify this, we simulated the images from single muons at $\theta_z = 87^\circ$ for different heights in the atmosphere and summed them up with the weights corresponding to the particle density at the appropriate height, as shown in Fig.~\ref{fig:profile_particles}.  Fig.~\ref{fig:simulated_proton_image} shows the result of stacking the individual muon signals which has a several degree long  tail morphology.

\subsection{Imaging of the nearly horizontal showers}

\begin{figure}
\includegraphics[width=\linewidth]{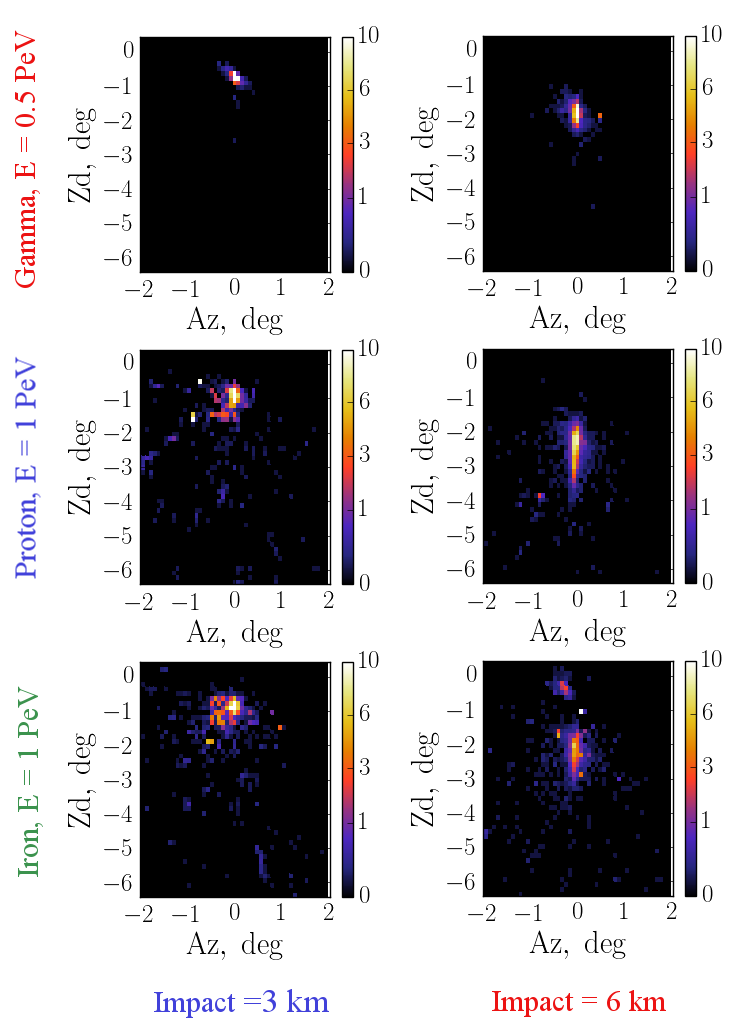}
\caption{Images and time profiles of 0.5 PeV \gr, 1 PeV proton and 1~PeV  iron induced EAS incident at $\theta_z=87^\circ$. Left column shows the images for the EAS impact parameter $d_{EAS}=3$~km, right panels are for $d_{EAS}=6$~km.}
\label{fig:images_cherenkov} 
\end{figure}

The overall morphology of the images of strongly inclined showers includes the compact electron component along with the  more extended muon component. This is shown in  Fig. \ref{fig:images_cherenkov}. The three rows of the figure show  the \gr, proton and iron EAS images. One could readily notice the difference between the \gr\ and proton / iron shower images. As discussed above, the \gr\ EAS image is more compact because it lacks the muon component. The muon component of the proton and iron EAS images appears as a ``halo'' (for small impact parameter showers shown in the left column of Fig. \ref{fig:images_cherenkov}) or as a ``tail'' (for larger impact parameter showers shown in the right column of the figure)  next to the compact image of the electron component.  

\begin{figure}
\includegraphics[width=\linewidth]{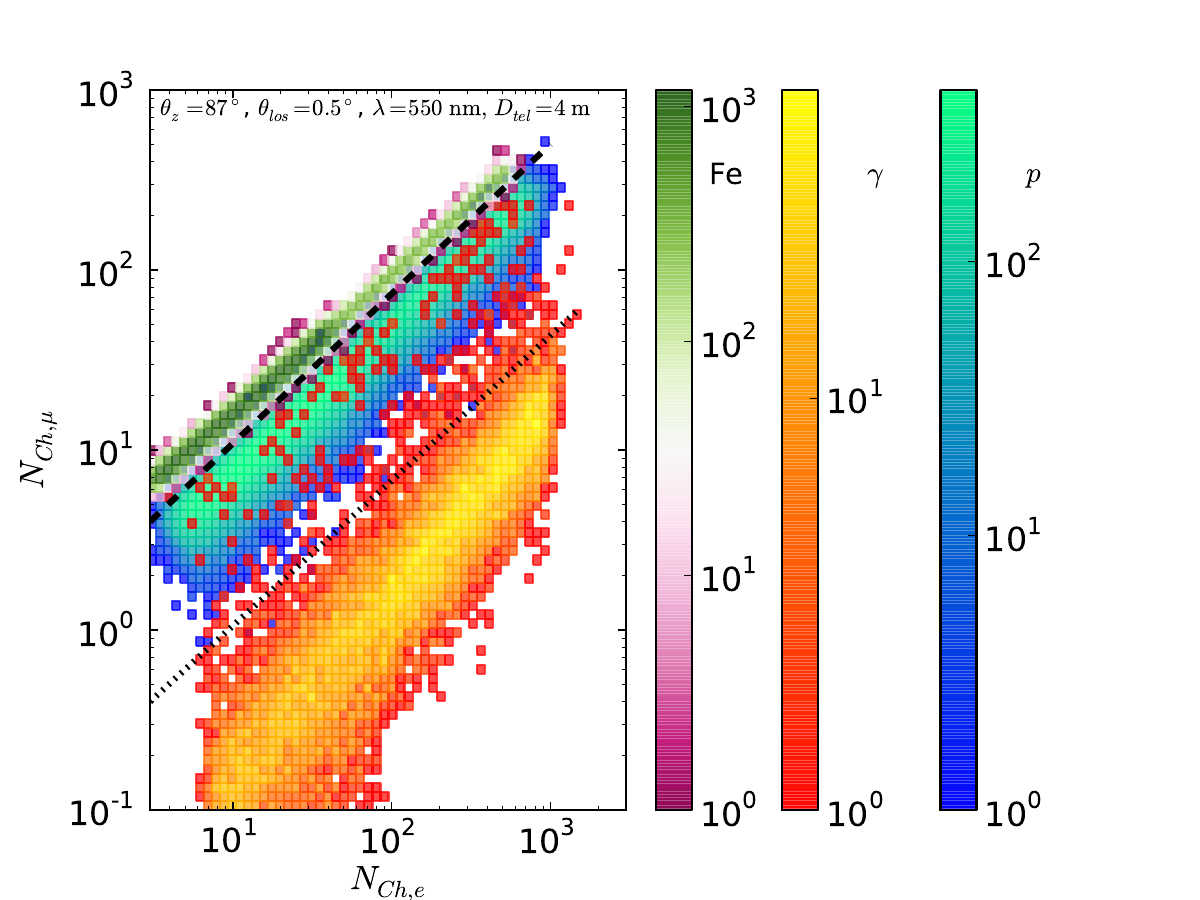}
\caption{Distribution of  the sizes of electron and muon components of the EAS images  for proton (blue-green), iron (green-magenta) and \gr\ (red-yellow) induced EAS with energies in the 1-100 PeV energy range incident at $\theta_z=87^\circ$ with $\theta_{los}=0.5^\circ$.  The inclined dashed and dotted line show the separation of iron / proton and proton+nuclei / \gr\ showers.}
\label{fig:scatter_plot} 
\end{figure}

The difference between the proton and iron EAS image is more subtle: it is in the relative normalisation of the electron  and muon components of the image. This is illustrated by Fig. \ref{fig:scatter_plot} which shows a scatter plot of the sizes $N_{Ch,\mu},\  N_{Ch,e}$ (number of photoelectrons) of the muon and electron components of the images of \gr, proton and iron showers.
One could see that the size of the muon component of the image is systematically higher for the iron showers, compared to the proton showers. The iron EAS are also characterised by a tight relation between  $N_{Ch,\mu}$ and $N_{Ch,e}$, while proton induced EAS have a larger spread of the $N_{Ch,\mu},\  N_{Ch,e}$ values. Still larger spread of the statistics of the muon component of the image and its overall much lower normalisation is evident for the \gr\ induced EAS images.

\subsection{Timing properties of the images}

Conventional imaging techniques of IACT rely on cameras with extremely fast readout electronics sampling the EAS signal on nanosecond time scales. These time scales are characteristic for the EAS incident at low zenith angles. Cherenkov photons arriving from different altitudes from a shower misaligned with the line of sight by an angle $\theta_{los}\simeq \alpha_{Ch}$ ($\alpha_{Ch}\simeq 1^\circ$ is the Cherenkov angle in the atmosphere) appear nearly simultaneously in the telescope camera. 

\begin{figure}
\includegraphics[width=\linewidth]{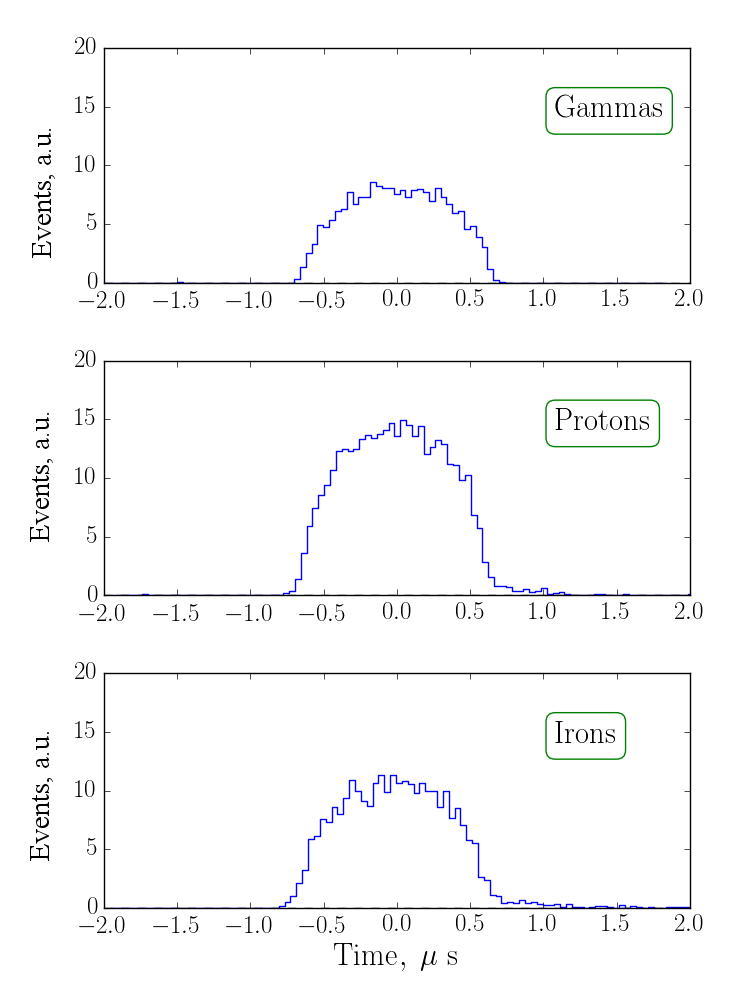}
\caption{Time profiles of formation of the images of the 0.5 PeV \gr, 1 PeV proton and 1PeV iron EAS incident at zenith angle $\theta_z=87^\circ$ with the impact parameter of 3~km. The zero moment of time for proton and iron is $2906~\mu s$ from the moment of the first interaction in the atmosphere. The arrival of the gamma shower is additionally delayed by $24~\mu s$, as its electron/muon components develop faster; protons and iron first interact though other channels and due to their superluminal motion develop the electromagnetic part closer to the observer.}
\label{fig:time_profile} 
\end{figure}

This fast image formation time is not applicable for the nearly horizontal showers because of the large spatial extent of the showers.  The characteristic time scale of the image could be estimated from the time delay between the time of arrival of light and particles (muons) of the EAS. Taking a shower perfectly aligned along the line of sight (i.e. a shower with zero impact parameter) one could estimate the time spread of the signal in the following way. The light travels with the speed $v=c/n$, so that the time difference accumulated over the propagation distance $\sim r_{horizontal}$ is
\begin{equation}
\Delta t\sim (n-1)r_{horizontal}\simeq 300\mbox{ ns}
\end{equation}
The image formation time scale is still longer for the showers misaligned with the direction to the telescope. The time profiles of the signal from nearly horizontal EAS are shown  in  Fig.~\ref{fig:time_profile}. One could see that the typical time scales do not strongly depend on the particle type. The signal is spread over $\Delta t\sim 1\ \mu$s for the \gr, proton and iron showers.  

\subsection{Imaging of the showers at the zenith angles close to $70^\circ$}

\begin{figure}
\includegraphics[width=1.0\linewidth]{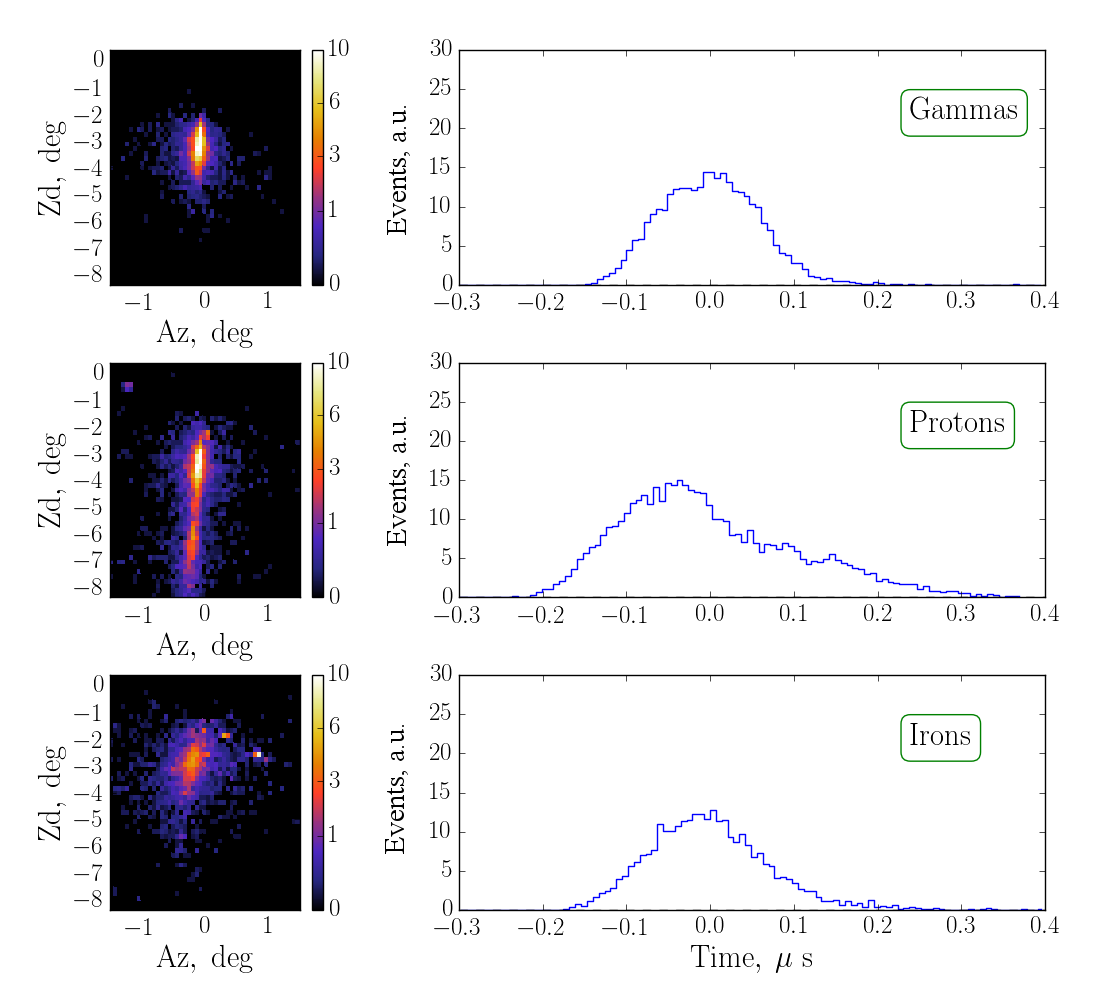}
\caption{Images (left column) and time profiles (right column) of the EAS initiated by 50 TeV \gr, 100 TeV proton and 100 TeV iron incident at $\theta_z=70^\circ$ with the impact parameter of 2 km. The time axis for all three cases is shitfted by $1038.2 \mu s$.}
\label{fig:images_zd70} 
\end{figure}

The method of observation of the muon halo (or tail) component of the EAS images works also for strongly  inclined, rather than nearly horizontal, EAS. An advantage of observations at more moderate zenith angles is the lower energy threshold. Such EAS develop closer to the telescope (at the distance $H_{atm}/\cos(\theta_z)$, which is about 30~km for $\theta_z=70^\circ$) and produce more light.

However, detection of the muon component of moderately inclined showers is more challenging 
because of the shorter path length available for the muons. This leads to a smaller amount of Cherenkov light in the muon component of the shower. From Fig.~\ref{fig:profiles_thetaz} one could see that EAS produced by 1~PeV protons do not possess a muon ``plateau'' in the longitudinal profile at all, if incident at zenith angles $\theta_z\lesssim 60^\circ$. 

Starting from $\theta_z\simeq 70^\circ$ the muon component of the EAS starts to be identifiable in the longitudinal profiles of PeV energy scale EAS because the electron component fades away high in the atmosphere. Fig.~\ref{fig:images_zd70} shows the images and time profiles of EAS initiated by the 50~TeV \gr s and 100~TeV protons and iron nuclei. One could see the same qualitative difference of the longitudinal profiles and images of the \gr\ and proton / iron EAS: the \gr\ shower images lack the muon tails. These muon tails are  hardly visible within the $\sim 5^\circ$ distance from the nominal EAS arrival direction. Observation of the muon component of distant EAS incident at zenith angles about $70^\circ$ requires the telescopes with very large FoVs, $\Theta_{FoV}\gtrsim 10^\circ$.  

Smaller distance scale of the showers incident at moderate zenith angles leads to their shorter time spread, as one could see from the right column of Fig.~\ref{fig:images_zd70}. The typical time scales of \gr, proton and iron EAS are $\sim 100$~ns in this case (compare with $\sim 1\mu$s time scale of nearly horizontal EAS).

\section{Discussion}

We have shown that observations of strongly inclined EAS with a system of wide FoV IACTs overlooking the horizon strip could be used to characterise the muon and electron content of the EAS and in this way to determine the identity of the primary high-energy particle which initiated it.

\subsection{Suppression of the cosmic ray background for \gr\ observations}

The possibility for particle identity determination could be used to reject the proton and nuclei induced cosmic ray background in \gr\ observations at the highest energies ($E\gtrsim 100$ TeV energy band).  Large number of Galactic \gr\ sources from the HESS Galactic Plane survey \citep{HESS_survey} have their spectra extending to the highest energies (tens of TeV) without a cutoff. Exploration of the 100~TeV band is important for understanding of the nature of particle accelerators operating in these sources.  KACADE limits on the flux from the Northern sky in this energy band  \cite{kascade} make it clear that observations of astronomical sources in the 100 TeV band would require very efficient background suppression.   

\begin{figure}
\includegraphics[width=\linewidth]{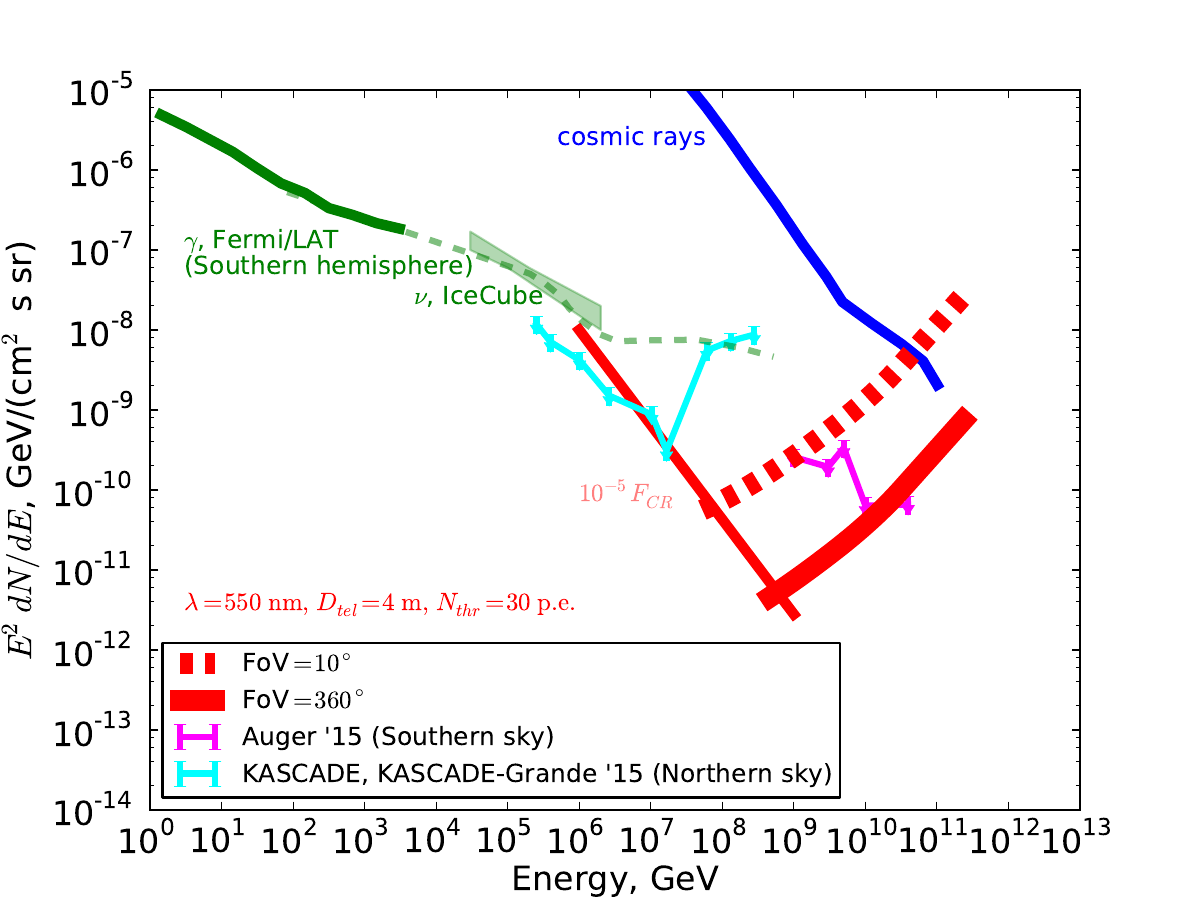}
\caption{Sensitivity limit of  Cherenkov telescopes with parameters specified in the text in the horizontal \gr\ shower observation mode. Thick red solid, and dashed curves show the differential sensitivity corresponding to one event per energy decade calculated assuming that the FoV spans the entire $360^\circ$ strip along the horizon. Thick dashed line corresponds to the $10^\circ$ FoV. Thin red solid line shows the flux level corresponding to the $10^{-5}$ of the cosmic ray flux. \textbf{Green dashed line shows powerlaw extrapolation of the Fermi/LAT spectrum modified by absorption on cosmic microwave backgorund over the distance scale $8$~kpc. Green shaded range shows IceCube measurement of astrophysical neutrino flux. }}
\label{fig:sensitivity} 
\end{figure}

An analytic approximate expression for the sensitivity of  an optimised IACT system overlooking the entire $360^\circ$ strip along the horizon  could be found in the following way. The horizontal looking IACT could observe nearly horizontal EAS events  over the area
\begin{equation}
\label{eq:aeff1}
A\sim 2\pi r_{horizontal} H_{atm}\simeq 2\times 10^{4}\mbox{ km}^2
\end{equation}
The showers are observable within a solid angle
\begin{equation}
\label{eq:omega}
\Omega\sim \pi (\alpha_{Ch}+\gamma_{Ch}^{-1})^2\simeq 6\times 10^{-3}
\end{equation}
The minimal detectable flux of diffuse all-sky  \gr\ background  in an exposure $T_{exp}\sim 1$~yr with the observation duty cycle $\kappa\simeq 0.1$ is
\begin{equation}
F_{min}\simeq \frac{E}{A\Omega \kappa T_{exp}}\simeq 10^{-11}\left[\frac{E}{10^8\mbox{ GeV}}\right]\frac{\mbox{GeV}}{\mbox{ cm}^2\mbox{ s sr}}
\end{equation}
Observations from high altitude $H\gg H_{atm}$, e.g. with a balloon or space borne system of Cherenkov telescopes like CHANT \cite{chant} will have somewhat better sensitivity at the highest energies, due to the larger effective area, estimated as 
\begin{equation}
\label{eq:aeff2}
A_{CHANT}\sim 2\pi \sqrt{2 H R_\oplus}H_{atm}\simeq 10^{5}\left[\frac{H}{300\mbox{ km}}\right]^{1/2}\mbox{ km}^2
\end{equation}

A more precise estimate of sensitivity is shown in Fig. \ref{fig:sensitivity}.  The sensitivity curves correspond to the flux level which produces  one detectable event per decade of energy  per one year of observations with duty cycle $0.1$ \cite{chant}.   The computation is done using the numerical code introduced in the Ref. \cite{chant}. The efficiency of the telescope optical system (product of optical system and photon detection efficiency of photosensors)  is assumed to be $\epsilon=0.2$. The telescope modules are assumed to have the aperture  $D_{tel}=4$~m. The calculation  takes into account Rayleigh scattering of the Cherenkov photons in the atmosphere for the reference  wavelength $\lambda=550$~nm. 

The background for the \gr\ signal is the flux of EAS initiated by  multi-PeV protons and heavier nuclei. Blue curve in Fig. \ref{fig:sensitivity} shows the level of the cosmic ray flux, while the red thin solid line shows the level of residual cosmic ray background estimated assuming the suppression of the cosmic ray flux by a factor $10^5$.  In the energy range below several PeV such suppression factor is reached by the KASCADE array \cite{kascade}.  

From  Fig. \ref{fig:scatter_plot} one could see that a comparable suppression of the proton / nuclei cosmic ray backgorund could be achieved also with the IACTs in the near horizontal shower observation mode. The statistics of proton / iron and \gr\ events in the figure is  $7\times 10^4$ iron EAS events, $7\times 10^4$ proton events and $10^4$ \gr\ events. Imposing a suitable restriction on the strength of the muon component (the ratio of the muon to electron image component sizes above the dotted line), observations of highly inclined EAS with an IACT system could achieve the $\sim 10^5$ efficiency of rejection of the proton and nuclei induced EAS, while rejecting only a small fraction (about $\lesssim 10^{-3}$) of gamma showers which occasionally have a sizeable muon component.

Such a "muon veto" method of suppression of the proton / nuclei background works only if the muon component of the EAS image is detectable on top of the night sky background.  The rate of the dark  night sky background is estimated as \cite{chant}
\begin{eqnarray}
{\cal R}_{nsb} & \sim  & 10^8\left[\frac{\epsilon}{0.1}\right]\left[\frac{D_{\rm tel}}{4\mbox{ m}}\right]^2\left[\frac{\Omega}{0.2\mbox{ sq.deg.}}\right]\left[\frac{\Delta\lambda}{100\mbox{ nm}}\right] \nonumber \\
& \times & \left[\frac{\lambda}{400\mbox{ nm}}\right]\frac{1}{\mbox{ s}} \,,
\label{eq:back}
\end{eqnarray}
where $\Omega$ is the solid angle span by the image of the electron or muon component of the EAS, $\lambda$ and $\Delta\lambda$ are the central wavelength and the width of the wavelength window of the photosensors of the telescope camera. The actual night sky backgorund rate is increasing with zenith angle but Eq. (\ref{eq:back}) still provides a good estimate even at large zenith angles. The statistics of the night sky background within the EAS detection trigger  time window $\Delta t\sim 100$~ns (for the observations at $\theta_z\sim 70^\circ$, see Fig. \ref{fig:images_zd70})   at the level of $N_{nsb}=\Delta t {\cal R}_{nsb}\simeq 10$ for a shower spanning a solid angle $\Omega\simeq 0.2\times 1$~sq.deg. The muon halo is detectable at $\gtrsim 3\sigma$ level as soon as its statistics becomes comparable to that of the night sky background,  $N_{thr}\sim 10$ and at $6\sigma$ level for the photon statistics twice as much.  From Fig. \ref{fig:images_zd70} one could see that for a 4~m diameter telescope such signal statistics corresponds to the threshold in the 10~TeV range. 

Cosmic ray background-free \gr\ observations in the $E>100$~TeV energy range will explore the population of Galactic sources capable to accelerate particles to the energies beyond PeV.  0.1-10~PeV  \gr s from other galaxies could not reach the Earth because of absorption in interactions with the photons of Cosmic Microwave Background (CMB). The mean free path of the PeV \gr s through the CMB is estimated as 
\begin{equation}
\lambda_{\gamma\gamma}=\frac{1}{\sigma_{\gamma\gamma}n_{CMB}}\simeq 8\mbox{ kpc}
\end{equation} 
where $\sigma_{\gamma\gamma}\simeq 10^{-25}$~cm$^{-2}$ is the maximal value of the pair production cross-section at the energy four times larger than the threshold
\begin{equation}
E_\gamma=\frac{4m_e^2}{\epsilon_{CMB}}\simeq 10^{15}\left[\frac{\epsilon_{CMB}}{10^{-3}\mbox{ eV}}\right]\mbox{ eV}
\end{equation}
with $\epsilon_{CMB}\simeq 10^{-3}$~eV being the average energy of the CMB photons \cite{aharonian_book}. The mean free path is about the distance from the Sun to the Galactic Centre. Thus, only emission from sources in about a half of the Milky Way, including the Galactic Centre "PeVatron" \cite{pevatron}  could reach the Earth. 

Apart from isolated Galactic PeVatrons, strong enough diffuse \gr\ emission in this energy band could be generated by interactions of cosmic rays in the interstellar medium. The Galactic diffuse emission flux level in the PeV range is uncertain. An estimate  could be found via extrapolation of the Milky Way  spectrum measured by Fermi Large Area Telescope (LAT) in the energy band below 10~TeV \cite{neronov16} shown by the green curve in Fig. \ref{fig:sensitivity}. This estimate follows a powerlaw with the slope fixed by the Fermi/LAT data. An absorption due to the $\gamma\gamma$ pair produciton on CMB appears as a deep in the spectrum in the PeV energy range. It is useful to note that the high-energy extrapolation of the Fermi/LAT \gr\ spectrum is consistent with the measurement of diffuse astrophysical neutrino flux by IceCube telescope \cite{IceCube_PeV,IceCube_3yr,IceCube_ICRC}. This is expected if the astrophysical neutrino flux has a sizeable Galactic component \cite{neronov14a,neronov14,neronov15,neronov16}. 

Extension of the diffuse emission spectrum in the multi-PeV range is possible if the Galaxy hosts cosmic ray accelerators able to produce cosmic rays with energies up to 100~PeV and higher. Otherwise, the diffuse Galactic \gr\ and neutrino spectra should have high-energy cut-off. Combined \gr\ and neutrino measurements will provide a precise characterisation of the Galactic diffuse emission and measurement of such a cut-off. 

The extragalactic sky "opens" again at the energies much above PeV due to the decrease of the pair production cross-section much above the threshold. The mean free path of the 10 EeV energy \gr s is several Mpc.  This long mean free path opens a possibility of detection of \gr\ emission from the nearest extragalactic UHECR accelerator(s) (for example, Centaurus A radio galaxy \cite{cena}) if they are present within this distance limit. 

\subsection{Cosmic ray flux composition measurements}

Measurement of the muon content of the EAS is actively used for characterization of the cosmic ray flux composition by the surface EAS arrays, like KASCADE \cite{kascade} or future Auger Prime detector \cite{augerprime}. Surface arrays directly measure the flux of electrons and muons at the ground level.  The method described in the previous sections provides a different type of measurement of the muon content of the EAS via characterisation of the overall Cherenkov photon statistics in the electron and muon  components of the EAS image in an IACT telescope camera. This measurement is  integrated over a range of altitudes. In this sense, the  method of the measurement of EAS electron and muon content discussed above is complementary to that employed by the surface detector arrays. 

Fluorescent detectors extract information on the cosmic ray particle identity from the measurement of the  depth of the shower maximum, the technique used by the Pierre Auger Observatory  \cite{auger_composition,auger_composition1} and  Telescope Aarray experiments \cite{ta_composition}. The depth of the shower maximum could also be reconstructed from the lateral distribution of the Cherenkov light from the EAS  as it is done in TUNKA experiment \cite{tunka}. From Fig.~\ref{fig:profile_cherenkov} one could see that the observations of strongly inclined showers also allow a measurement of the depth of the maxima of the electron components of the showers. From these figures one could judge the difference in the depth of the maxima of proton and heavier nuclei induced EAS. Measurements of the shower maxima depths with IACT systems provide an additional observable shower characteristics sensitive to the  primary particle identity.  

From Fig.~\ref{fig:profile_cherenkov} it is clear that also combined measurement of the muon content and of the depth of the shower maximum is possible for the observations of strongly inclined  EAS with IACT systems. This provides a potentially more powerful constraint on the identity of the primary cosmic ray particles, compared to the measurements of only the muon flux on the ground or of the depth of the EAS maximum. 

Muon detectors of surface EAS arrays are limited in size. For example, measurements of the muon content of the EAS in KASCADE-Grande were possible only within the KASCADE array of the area $A=0.2\mbox{ km }\times 0.2\mbox{ km}=0.04$~km$^2$. KASCADE has collected cosmic rays from a large solid angle $\Omega\sim 1$~sr, so that the grasp of the system was $A\Omega\sim 0.04$~km$^2$sr. From Eqs.  \ref{eq:aeff1}, \ref{eq:omega} one could derive a duty-cycle corrected  grasp of IACT system observing at high zenith angle $\kappa A\Omega\gtrsim 10$~km$^2$sr which is several orders of magnitude larger. This potentially provides higher signal statistics and a broader observation energy range.  From Fig. \ref{fig:sensitivity} one could see that a year long exposure with an IACT system providing a $360^\circ$ wide field-of-view along the horizon will detect events with energies up to the UHECR range.  

A series of composition changes in the cosmic ray spectrum occurs in the energy range from $10^{15}$~eV to $10^{20}$~eV \cite{kascade}. Numerous EAS array  experiment have reported composition changes from light to heavy nuclei in the knee energy range and then back to the light in the energy range above $10^{17}$~eV.  Pierre Auger Observatory observes a light-to-heavy composition change in the energy decade between $10^{19}$~eV and $10^{20}$~eV \cite{auger_composition,auger_composition1}. The detailed shapes of the spectra of light and heavy nuclei across the reported change energies  are dependent on the phenomenological models of hadronic interactions used in the calculation of the EAS. Combined measurements of the muon content and of the depth of the shower maximum with IACTs would provide additional constraints on the hadronic interaction models and reduce uncertainties in the calculation of details of the composition changes. 

\section{Conclusions}

Observations of strongly inclined EAS with IACT systems open a range of possibilities for cosmic ray background free \gr\ astronomy at the highest energy end of the astronomical observations window (above 100~TeV) and provide a possibility for the measurement of composition of the cosmic ray flux in the energy range up to UHECR. This is possible because a sizable contribution to the shower signal is provided by muons. Muon signal forms a distinct halo or tail component of the EAS image. This component is readily distinguishable from the electromagnetic component of the image. 

The IACT setup optimized for the high zenith angle observations is different from the conventionally chosen setup optimized for low zenith angles. In particular, the telescope cameras have to be equipped with photosensors sensitive in the visible, rather than UV, wavelength range.  The readout of the telescope camera should be able to sample the shower signal on very long time scales (0.1-1~$\mu$s, for zenith angles between $70^\circ$ and $90^\circ$, compared to the conventional 1-10~ns time scale of low zenith angle observations). Large angular extent of the muon component of the shower images, low overall count rate and high energy threshold of high zenith angle observations favours the use of the optical systems with very wide FoV (much larger than $\lesssim 10^\circ$ FoV of the currently operational IACTs and of the next generation facility CTA).   

\bibliography{Muon_showers}

\end{document}